\documentclass[showpacs,aps,twocolumn]{revtex4}
\usepackage{epsfig}
\usepackage{graphicx}
\usepackage{amsmath,amssymb,amsfonts}
\usepackage{array}
\usepackage{url}
\usepackage{multirow}
\usepackage{float}
\usepackage{lineno}
\usepackage{xspace}
\usepackage{ulem}

\def\ba{\begin{eqnarray}}
\def\ea{\end{eqnarray}}

\usepackage[usenames,dvipsnames]{color}
\definecolor{darkblue}{RGB}{0,0,196}
\usepackage[colorlinks=true,linkcolor=darkblue,citecolor=darkblue,urlcolor=darkblue]{hyperref}

\begin{document}
\title{Conserved-charge diffusion in hot and dense QCD matter}

\author{Dushmanta Sahu}
\email{Dushmanta.Sahu@cern.ch}
\affiliation{Instituto de Ciencias Nucleares, UNAM, Apartado Postal 70-543, Coyoacán, 04510, México City, México}
\author{Kangkan Goswami}
\email{Kangkan.Goswami@cern.ch}
\author{Raghunath Sahoo}
\email{Raghunath.Sahoo@cern.ch}
\affiliation{Department of Physics, Indian Institute of Technology Indore, Simrol, Indore 453552, India}

\date{\today}
\begin{abstract}
We study the coupled diffusion of baryon number, electric charge, and strangeness in hot and dense quark matter using a three-flavor Nambu–Jona-Lasinio model with self-consistently generated constituent quark masses. The complete diffusion matrix is evaluated within a relaxation-time kinetic approach, including its diagonal and off-diagonal components. We find that the off-diagonal transport coefficients exhibit distinct temperature and baryon-chemical-potential dependence associated with the flavor-dependent composition of the charge carriers. We further investigate generalized charge diffusivities and mobilities obtained from the corresponding susceptibilities. At finite baryon chemical potential, the baryon and strangeness mobility scales cross, reflecting the different density dependence of their transport coefficients and susceptibilities. The results provide model benchmarks for coupled conserved-charge transport in baryon-rich QCD matter.
\end{abstract}
\maketitle

\section{Introduction}

Ultra-relativistic hadronic and nuclear collisions at the Large Hadron Collider (LHC) at CERN and the Relativistic Heavy Ion Collider (RHIC) at BNL help us explore the behavior of hot and dense QCD matter. These experiments have been key to revealing the unique properties of the short-lived exotic state of matter known as quark-gluon plasma (QGP), which is a deconfined and locally thermalized state of quarks and gluons~\cite{PHENIX:2004vcz, PHOBOS:2004zne, STAR:2005gfr}. In recent times, one of the most surprising discoveries is that the QGP is a perfect fluid, having an incredibly low ratio of shear viscosity to entropy density ($\eta/s$), among the lowest ever observed in nature~\cite{Romatschke:2007mq, Heinz:2013th}. Significant progress has also been made in understanding bulk viscosity, $\zeta/s$, which is expected to exhibit enhanced behavior in the vicinity of the QCD transition~\cite{Karsch:2007jc, Meyer:2007dy}. More recently, increasing attention has been devoted to the scaled electrical conductivity ($\sigma_{\rm el}/T$), which plays an important role in charge transport and the evolution of electromagnetic fields in heavy-ion collisions. It has been investigated using complementary approaches, including lattice QCD, perturbative QCD, kinetic theory, and effective models~\cite{Aarts:2020dda, Arnold:2000dr, Mitra:2016zdw}.\\

However, an important aspect of QCD transport that remains comparatively unexplored is the diffusion of conserved charges. In relativistic heavy-ion collisions, the transport processes involve the coupled diffusion of conserved charges, including baryon number, electric charge, and strangeness, which can be described by a diffusion matrix with both diagonal and off-diagonal components~\cite{Greif:2017byw}. The importance of such diffusion processes increases at lower collision energies, where the baryon chemical potential becomes substantially larger. This makes the RHIC Beam Energy Scan (BES) program particularly relevant, as it explores QCD matter over a broad range of temperature and baryon chemical potential. Hydrodynamic studies incorporating baryon diffusion have demonstrated the relevance of finite-density transport effects for the dynamical evolution of matter produced at collision energies accessible to the BES program~\cite{Denicol:2018wdp}. Moreover, microscopic calculations have shown that the diffusion coefficients exhibit a nontrivial dependence on both temperature and baryon chemical potential~\cite{Fotakis:2021diq}. Motivated by these considerations, in the present work, we calculate the diffusion matrix and its associated transport coefficients over the $(T,\mu_B)$ plane and thereby target the finite-density regime explored by the RHIC BES and the FAIR CBM experiments. This shift in experimental focus opens new doors for understanding how conserved charges flow and spread in extreme states of matter, a puzzle that could reveal deeper insights into the fundamental nature of the strong force. Conserved-charge transport is also becoming accessible experimentally at small $\mu_B$. STAR collaboration has recently compared baryon and electric-charge transport directly and finds a baryon-to-charge transport ratio of $\approx1.8$ at midrapidity in central isobar collisions, well above the value of unity expected if both charges were carried by valence quarks~\cite{STAR:2024lvy}. This makes the relative transport of $B$ and $Q$, which is encoded in the off-diagonal elements of the diffusion matrix, an experimentally relevant question. \\

The diffusion matrix for the three conserved charges is given by~\cite{Greif:2017byw},
\begin{equation}
\begin{pmatrix}
j_{\rm B}^{\mu} \\
j_{\rm Q}^{\mu} \\
j_{\rm S}^{\mu}
\end{pmatrix}
=
\begin{pmatrix}
\kappa_{\rm BB} & \kappa_{\rm BQ} & \kappa_{\rm BS} \\
\kappa_{\rm QB} & \kappa_{\rm QQ} & \kappa_{\rm QS} \\
\kappa_{\rm SB} & \kappa_{\rm SQ} & \kappa_{\rm SS}
\end{pmatrix}
\begin{pmatrix}
\Delta^{\mu} \alpha_{\rm B} \\
\Delta^{\mu} \alpha_{\rm Q} \\
\Delta^{\mu} \alpha_{\rm S}
\end{pmatrix},
\end{equation}
where $\alpha_q=\mu_q/T$ denotes the thermal potential associated with the conserved charge $q$, with $q\in\{B,Q,S\}$, $B$, $Q$ and $S$ are baryon number, electric charge and strangeness respectively. The off-diagonal coefficient $\kappa_{qq'}$ characterizes how a gradient in thermal potential of one conserved charge, $q'$, can drive a diffusion current of another conserved charge $q$. The diagonal elements describe the conventional diffusion of individual conserved charges, whereas the off-diagonal elements encode cross-diffusion between different charge sectors. Thus, for example, a strangeness gradient can contribute to the baryon current through a nonzero $\kappa_{\rm BS}$. Such couplings arise naturally because the microscopic degrees of freedom can carry more than one conserved quantum number simultaneously, and their magnitude depends on the charge content and microscopic dynamics of the medium~\cite{Fotakis:2021diq, Fotakis:2019nbq}. Consequently, the off-diagonal diffusion coefficients provide additional information on the interplay among the conserved charges that is not contained in the corresponding diagonal transport coefficients. In an interacting medium with dynamically generated and flavor-dependent quark masses, these coefficients can also be sensitive to flavor-symmetry breaking, providing a transport-level complement to thermodynamic observables such as the chiral condensate.\\

To establish a direct connection between the diagonal elements of the diffusion matrix and the transport of the individual conserved charges, we consider the diagonal diffusion matrix coefficients $\kappa_{qq}$ and susceptibilities $\chi_{q}$ corresponding to the baryon number, electric charge, and strangeness sectors. Here, the diagonal susceptibilities are estimated as $\chi_q = \frac{\partial n_q}{\partial \mu_q} = \frac{\partial^2 P}{\partial \mu_q^2}$. For each conserved charge, the associated diagonal diffusivity can be defined through the corresponding charge susceptibility as,
\begin{equation}
D_q=\frac{\kappa_{qq}}{T~\chi_{q}}.
\label{eq2}
\end{equation}
This relation allows the transport coefficients $\kappa_{BB}$, $\kappa_{QQ}$, and $\kappa_{SS}$ to be expressed in terms of the corresponding baryon, electric-charge, and strangeness diffusivities. The diffusivities can subsequently be connected to the generalized conserved-charge mobility ($M_{q}$) through the Einstein--Smoluchowski relation,
\begin{equation}
M_q=\frac{D_q}{T}.
\end{equation}
These relations provide a direct link between the microscopic diffusion coefficients obtained from the transport calculation and the mobility of the corresponding conserved charges. The diffusivities and corresponding conserved-charge mobilities provide complementary measures of how efficiently the different conserved charges propagate through the strongly interacting medium. Their temperature and baryon-chemical-potential dependence can reveal changes in the microscopic charge carriers and transport dynamics, particularly in the vicinity of the QCD crossover, where critical fluctuations and rapidly varying susceptibilities may significantly modify charge transport.\\

The diffusion matrix in the context of heavy-ion collisions has been explored recently in the literature. It was first proposed by Greif \textit{et al.} in Ref.~\cite{Greif:2017byw}, where the authors explored the diffusion matrix in a hadron gas and a massless QGP using relativistic kinetic theory. It has also been explored within the dynamical quasiparticle model (DQPM) in Ref.~\cite{Fotakis:2021diq}, where the parton spectral functions and interaction rates are constrained by lattice thermodynamics. The cross-conductivities associated with the coupling between electric, baryon, and strangeness currents have also been investigated in an interacting hadron gas using the Green--Kubo formalism within the SMASH transport framework~\cite{Rose:2020sjv}. In addition, the effect of repulsive interactions has been studied in the diffusion matrix~\cite{Das:2021bkz}. Moreover, in Ref.~\cite{Goswami:2024hfg}, we extended the study of the diffusion matrix by including the charm quantum number within the hadron resonance gas with van der Waals-like interactions. These studies collectively span a wide range of microscopic pictures of the medium, from a hadronic gas of interacting resonances to a quasiparticle description of the deconfined phase. However, to our knowledge, none of these approaches is based on a model with explicit, dynamically generated chiral symmetry breaking, which is expected to play a central role in shaping transport coefficients in the vicinity of the QCD crossover.\\

In this article, we estimate the diffusion matrix for the conserved charges within the three-flavor Nambu--Jona-Lasinio (NJL) model~\cite{Nambu1961}, using the standard $SU(3)_f$ parametrization of Ref.~\cite{Klevansky1992} with couplings fixed as in Ref.~\cite{Rehberg1996b}. The NJL model shares the global symmetries of QCD, including chiral symmetry and its spontaneous breaking, and generates dynamical, temperature- and density-dependent constituent quark masses through a set of coupled gap equations. This provides a complementary perspective to the frameworks discussed above. Rather than treating the medium as a gas of hadronic resonances or as a quasiparticle ensemble constrained to reproduce lattice thermodynamics, the NJL model allows the chiral dynamics that drives the QCD crossover to feed directly and self-consistently into the transport coefficients themselves. This also enables us to study their dependence on $\mu_B$, where lattice QCD calculations remain severely limited because of the fermion sign problem. Our results provide NJL-based benchmarks for the charge transport coefficients that can be compared and used in hydrodynamic simulations of heavy-ion collisions and to study the QCD phase diagram.\\

This article is organized as follows. Section \ref{formulation} gives the formulation of the NJL model and the diffusion matrix. Section~\ref{result} discusses the result and the consequences of our current study. Finally, in section~\ref{summary}, we summarize our work and give an outlook for future studies.\\

\section{Formulation}
\label{formulation}
\subsection{Nambu--Jona-Lasinio Model}

The Nambu--Jona-Lasinio (NJL) model was originally introduced by Y. Nambu and G. Jona-Lasinio in 1961 as a model of interacting fermions, with nucleons as the relevant degrees of freedom~\cite{Nambu1961}. Following the formulation of quantum chromodynamics (QCD), the model was subsequently adapted to describe quarks and antiquarks, with the gluonic degrees of freedom effectively integrated out. In the present work, we employ the three-flavor ($SU(3)_f$) version of the NJL model, whose Lagrangian is given by~\cite{Klevansky1992, Rehberg1996b, Buballa:2003qv},
\begin{equation}
  \begin{aligned}
    \mathcal{{L}}_{NJL} =&\bar{\psi} \left( i \partial\!\!\!\!/- m_0 \right) \psi\\
             &+ \ G \sum_{a=0}^8 \left[   \left( \bar{\psi}            \lambda^a \psi \right)^2
                                        + \left( \bar{\psi} i \gamma_5 \lambda^a \psi \right)^2 \right] \\
             &- \ K \left[   \mathrm{det} \bar{\psi} \left( 1 - \gamma_5 \right) \psi
                           + \mathrm{det} \bar{\psi} \left( 1 + \gamma_5 \right) \psi \right],
  \end{aligned}
  \label{NJL_lagrangian}
\end{equation}
where $\psi=(u,d,s)$ represents the three-flavor quark field, with $N_f=3$ and $N_c=3$. The matrices $\lambda^a$ $(a=0,\ldots,8)$ are the Gell-Mann matrices in flavor space, while $G$ and $K$ denote the four-quark and six-quark coupling constants, respectively. The first term in Eq.~\eqref{NJL_lagrangian} describes the free quark contribution with the bare mass $m_0$. The second term represents a chirally symmetric four-quark interaction, while the determinant term introduces an effective six-quark interaction that accounts for the $U_A(1)$ anomaly through the 't Hooft interaction.

The symmetry group of the NJL model is the same as that of  QCD,
\begin{equation}
  \text{U}(3)_L \otimes \text{U}(3)_R =
  \text{SU}_V(3) \otimes \text{SU}_A(3) \otimes \text{U}_V(1) \otimes \text{U}_A(1).
\end{equation}
While chiral symmetry is spontaneously broken in the vacuum and explicitly broken by the bare quark mass $m_0$ in Eq. \eqref{NJL_lagrangian}, the axial $U_A(1)$ symmetry is explicitly broken by the last term, corresponding to the 't Hooft flavor-mixing interaction.

\begin{table} [!t]
  \caption{Parameters for the NJL model from Ref. \cite{Rehberg1996b}.\label{paramNJL}}
  \begin{ruledtabular}
    \begin{tabular}{ c c c c c }
      $\Lambda$ & $G\Lambda^2$ & $K\Lambda^5$ & $m_{0q}$ & $m_{0s}$ \\
      \hline
      \noalign{\vskip 3pt}
      602.3 MeV & 1.835 & 12.36 & 5.5 MeV & 140.7 MeV \\
    \end{tabular}
  \end{ruledtabular}
\end{table}

The effective quark masses can be obtained from this Lagrangian. Within the mean-field approximation, they are given by,

\begin{equation}
  m_i = m_{0i} - 4 G \langle\langle \bar{\psi_i}\psi_i \rangle\rangle
               + 2 K \langle\langle \bar{\psi_j}\psi_j \rangle\rangle
                     \langle\langle \bar{\psi_k}\psi_k \rangle\rangle,
  \label{massNJL}
\end{equation}
with $i \ne j \ne k$ denoting the three possible flavors of quarks, and $\langle\langle \bar{\psi}\psi \rangle\rangle$ being the chiral condensate in the mean field limit,
\begin{equation}
  \langle\langle \bar{\psi}_i\psi_i \rangle\rangle
  = - 2 N_c \ \int_0^\Lambda \frac{d^3 p}{(2\pi)^3}
  \frac{m_i}{E_{i}} [1 - f_q - f_{\bar q}],
  \label{NJL_condensate}
\end{equation}
with $E_{i} = \sqrt{{p}^2 + M_{i}^{2}}$, $M_{i}$ is the constituent quark mass, $\Lambda$ is the 3-momentum cutoff, and the Fermi-Dirac distribution
\begin{equation}
  \begin{aligned}
    f_q (p,T,\mu)&= \{\exp[(E - \mu) / T] + 1\}^{-1},\\
    f_{\bar q}(p,T,\mu)&= \{\exp[(E + \mu) / T] + 1\}^{-1}.
  \end{aligned}
  \label{fermidis}
\end{equation}

The total grand potential in the mean-field approximation is given by
\begin{equation}
\Omega(T, \mu_u, \mu_d, \mu_s) = \Omega_{\text{cond}} + \Omega_{\text{vac}} + \Omega_{\text{med}},
\end{equation}
where each term is defined as follows,

\begin{equation}
\Omega_{\text{cond}} = 2G \left( \phi_u^2 + \phi_d^2 + \phi_s^2 \right) - 4K \, \phi_u \phi_d \phi_s,
\end{equation}
with \( \phi_i = \langle \bar{\psi}_i \psi_i \rangle \).

\begin{equation}
\Omega_{\text{vac}} = -2N_c \sum_{i=u,d,s} \int_{0}^{\Lambda} \frac{d^3p}{(2\pi)^3} \, E_{i},
\end{equation}

\begin{align}
\Omega_{\text{med}} = -2N_c \sum_{i=u,d,s} \int \frac{d^3p}{(2\pi)^3} \Bigg[ 
& T \ln\left(1 + e^{-(E_i - \mu_i)/T} \right) \notag \\
& + T \ln\left(1 + e^{-(E_i + \mu_i)/T} \right) 
\Bigg].
\end{align}

The pressure is defined as the negative of the grand potential,
\begin{equation}
P(T, \mu_u, \mu_d, \mu_s) = - \left ( \Omega(T, \mu_u, \mu_d, \mu_s) - \Omega(0, 0 ,0, 0) )\right .
\end{equation}

\subsection{Diffusion matrix}

The transport properties of the system are characterized by the energy-momentum tensor and conserved charge currents. The energy-momentum tensor contains both ideal and dissipative components,

\begin{equation}
\begin{aligned}
T^{\mu\nu}
={}&-P g^{\mu\nu}
+\omega u^\mu u^\nu
+\eta\left(D^\mu u^\nu+D^\nu u^\mu
+\frac{2}{3}\Delta^{\mu\nu}\theta\right)\\
&-\zeta\Delta^{\mu\nu}\theta .
\end{aligned}
\end{equation}

where $P$ is the pressure, $\omega = \epsilon + P$ the enthalpy, $\eta$ and $\zeta$ are shear and bulk viscosities respectively, and $\Delta^{\mu\nu} = g^{\mu\nu} - u^\mu u^\nu$ projects orthogonal to the fluid velocity $u^\mu$. Here $D \equiv u^\mu\partial_\mu$ is the comoving derivative, $D^\mu \equiv \partial^\mu - u^\mu D$ is the spatial gradient operator, and $\theta \equiv \partial\cdot u$ is the expansion scalar. The conserved currents for charge $q$ have a similar decomposition,

\begin{equation}
J_q^\mu = n_q u^\mu + \kappa_{qq'}D^\mu\alpha_{q'}, \quad \alpha_q \equiv \mu_q/T
\end{equation}

where $n_q$ is the net charge density and $\kappa_{qq'}$ the diffusion coefficients that couple different charge currents.

To estimate the diffusion coefficients, we employ kinetic theory through the
Chapman--Enskog expansion of the distribution function of species $i$,
\begin{equation}
  f_i = f_i^0\,(1+\Phi_i),
  \label{eq:CE}
\end{equation}
where $f_i^0$ is the equilibrium Fermi--Dirac distribution. To first order in gradients, $\Phi_i$ separates into a
part driven by gradients of $\alpha_q$ and a part driven by gradients
of the flow velocity,
\begin{equation}
  \Phi_i = -\sum_q B_i^{q}\,p_i^\mu D_\mu\alpha_q \;+\; \Phi_i^{\rm viscous},
  \label{eq:Phi}
\end{equation}
where $\Phi_i^{\rm viscous}$ collects the terms proportional to the shear tensor
$D^\mu u^\nu+D^\nu u^\mu-\tfrac{2}{3}\Delta^{\mu\nu}\theta$ and to the
expansion scalar $\theta$, these determine the shear
and bulk viscosities $\eta$ and $\zeta$. Being tensor and scalar structures,
they do not contribute to the vector currents $J_q^\mu$ at linear order, so
they are not needed for the diffusion matrix and are not considered further.
The relaxation-time fixes the coefficient of the diffusive part approximation of the Boltzmann equation in the Landau frame,
\begin{equation}
  B_i^{q}=\frac{\tau_i}{E_i}\left(q_i-\frac{n_qE_i}{\omega}\right),
  \label{eq:B}
\end{equation}
with $E_i=u\cdot p_i$ the energy in the local rest frame, $\tau_i$ the relaxation time, $n_q$ the net density of charge $q$ and $\omega=\epsilon+P$
the enthalpy density. The term $n_qE_i/\omega$ enforces Landau matching, so
that the correction carries no net energy flow. Solving the relativistic Boltzmann equation in the relaxation time approximation yields the explicit expression for the diffusion matrix as, 

\begin{equation}
\kappa_{qq'} = \sum_i \int \frac{d^3p_i}{(2\pi)^3} \frac{p_i^2}{3E_i} \left(q_i - \frac{n_q E_i}{\omega}\right) \frac{\tau_i}{E_i} \left(q'_i - \frac{n_{q'} E_i}{\omega}\right) f_i^0
\label{Eq:Dif_Matrix}
\end{equation}

A detailed derivation of Eq.~\ref{Eq:Dif_Matrix} is given in Ref.~\cite{Das:2021bkz, Goswami:2024hfg}. It is evident from the above equation that the diffusion matrix is symmetric under the interchange of the conserved charge indices \(q\leftrightarrow q'\). Therefore, the off-diagonal diffusion coefficients satisfy the Onsager reciprocity relation $\kappa_{qq'}=\kappa_{q'q}$. This symmetry is a manifestation of Onsager reciprocity and ensures that the diffusion matrix is symmetric, as expected for a system satisfying microscopic reversibility.

The relaxation time $\tau_i$ encodes the elastic scattering between the
quarks, given as
\begin{equation}
\tau_i^{-1} = \sum_j n_j \langle \sigma_{ij} v_{ij} \rangle,
\end{equation}
where the thermal average of the cross section is given
by~\cite{Gondolo:1990dk}
\begin{equation}
\begin{aligned}
\left\langle \sigma_{ij}v_{ij}\right\rangle
={}&
\frac{1}
{8T m_i^2m_j^2
K_2(m_i/T)K_2(m_j/T)}\\
&\times
\int_{(m_i+m_j)^2}^{\infty}
\frac{ds}{\sqrt{s}}\,
\sigma_{ij}(s)\,
\left[s-(m_i+m_j)^2\right] \\
&\times
\left[s-(m_i-m_j)^2\right]
K_1\left(\frac{\sqrt{s}}{T}\right),
\end{aligned}
\label{eq:sigmav}
\end{equation}
and $n_j$ number density of species $j$. Rather than evaluate the full set of eleven $s$-dependent elastic NJL cross sections computed in Ref.~\cite{Marty:2013ita} from explicit quark/meson exchange diagrams, we take $\sigma_{ij}(s)$ to be energy-independent but temperature-dependent,
\begin{equation}
\sigma_{ij}(s) \to \sigma_0(T) = \sigma_0(T_{\rm ref})\left(\frac{T_{\rm ref}}{T}\right)^2,
\label{eq:sigmaT}
\end{equation}
with the $T^{-2}$ scaling taken from the behavior Ref.~\cite{Marty:2013ita}
extracts for the same-flavor elastic channels above $T_c$, attributed there to the increase of the exchanged meson masses with temperature, which suppresses the resonant $s$-channel enhancement. The overall normalization is fixed directly from their computed $\sigma_{u u\to uu}(T,\sqrt{s})$ (their Fig.~3b), which remains below $\sigma_0(T_{\rm ref}{=}0.20~{\rm GeV})\simeq 1.5$~mb across the temperature and energy range they consider; this value is consistent with the $\lesssim 2.5$~mb bound quoted more broadly for elastic $qq$ scattering in NJL/PNJL calculations. Equation~(\ref{eq:sigmaT}) then gives $\sigma_0(T)$ comparable in magnitude to the elastic parton cross sections ($\sim 0.6$--$5$~mb) required in
transport-model studies of RHIC elliptic flow~\cite{Molnar:2001ux}.

\section{Results}
\label{result}
\begin{figure*}[ht!]
\includegraphics[width = 0.99\linewidth]{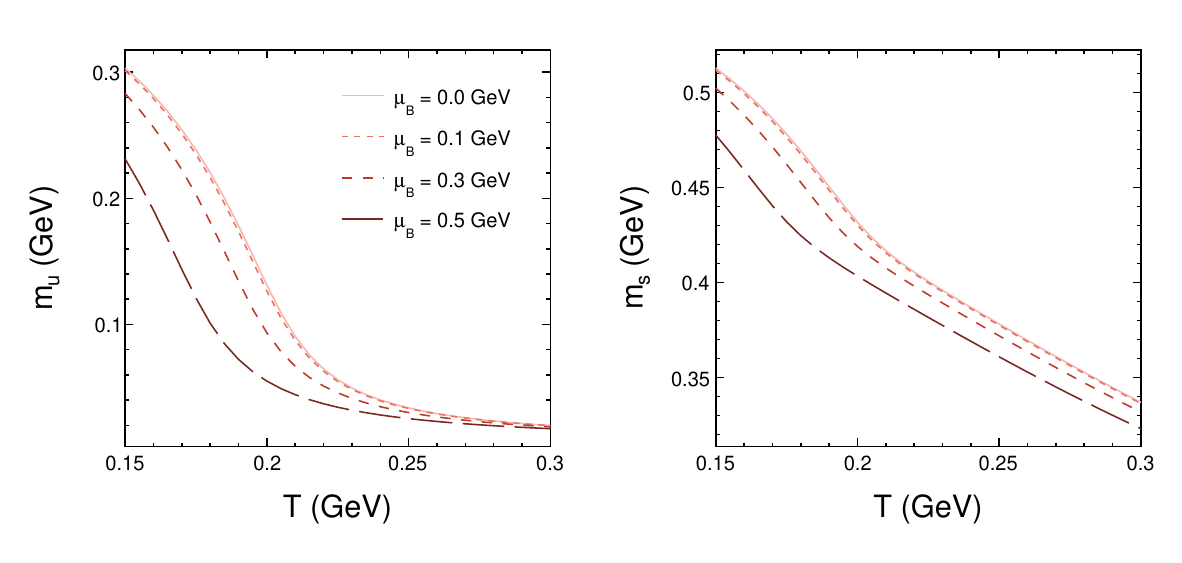}
    \caption{Dynamical mass of light quarks (left panel) and strange quark (right panel) as a function of temperature for different values of baryon chemical potential.}
    \label{Fig1}
\end{figure*}

\begin{figure*}[ht!]
\includegraphics[width = 0.99\linewidth]{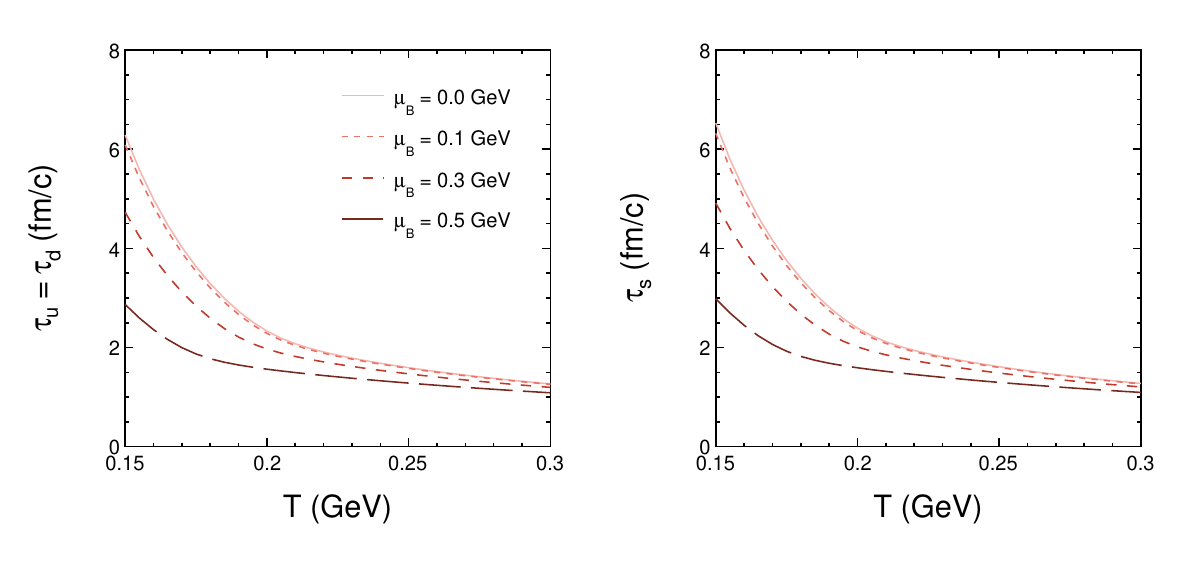}
    \caption{Relaxation time of light quarks (left panel) and strange quark (right panel) as a function of temperature for different values of baryon chemical potential.}
    \label{Fig2}
\end{figure*}

Figure~\ref{Fig1} shows the temperature dependence of the dynamically generated constituent quark masses. While both $m_u=m_d$ and $m_s$ decrease with temperature, the strange-quark mass remains substantially larger over the relevant range, reflecting persistent $\mathrm{SU}(3)_f$ breaking. This flavor-dependent mass splitting leads to different transport contributions from light and strange quarks and, consequently, leaves a characteristic imprint on the off-diagonal diffusion coefficients. Their temperature dependence therefore provides a complementary probe of flavor symmetry breaking in the medium.\\

\begin{figure*}[ht!]
    \centering
    \includegraphics[width = 0.99\linewidth]{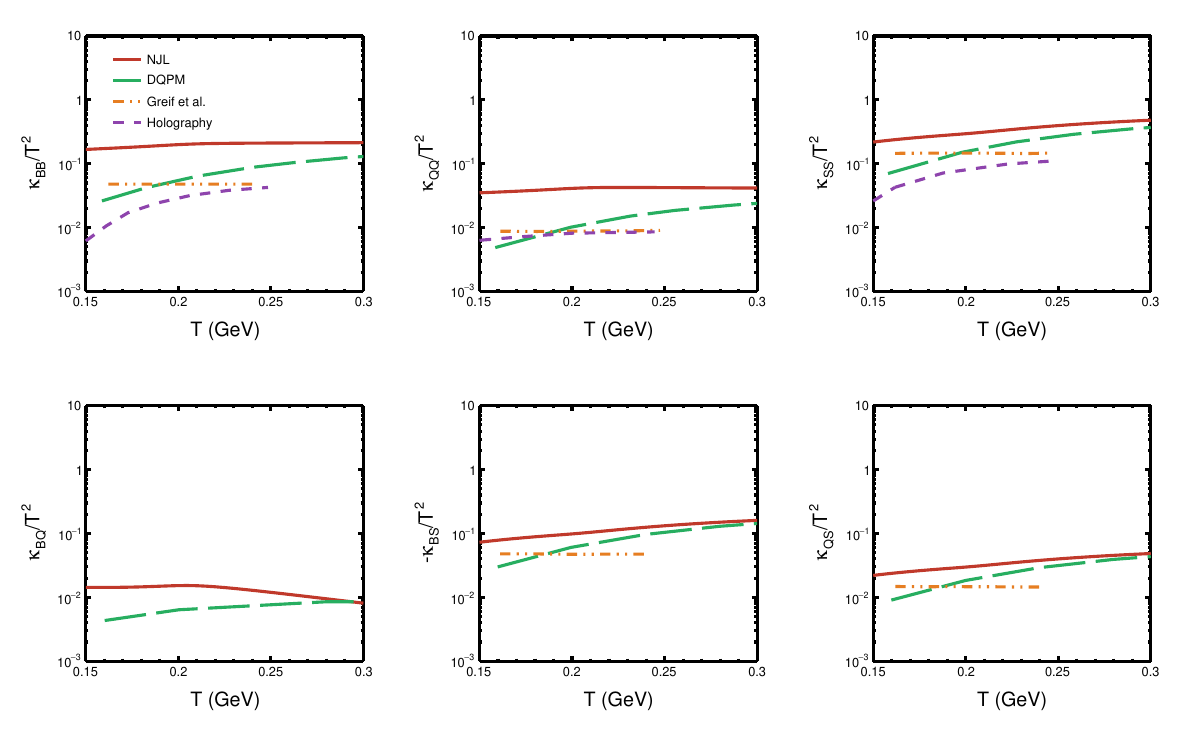}
    \caption{Diagonal and off-diagonal coefficients of the diffusion matrix as functions of temperature at zero baryon chemical potential. The NJL result is compared with results from DQPM~\cite{Fotakis:2021diq}, holography~\cite{Rougemont:2017tlu}, and kinetic theory~\cite{Greif:2017byw}.}
    \label{Fig3}
\end{figure*}
Figure~\ref{Fig2} shows the flavor-dependent quark relaxation times $\tau_u=\tau_d$ and $\tau_s$, obtained from the thermally averaged, screened quark-quark scattering rate. Both relaxation times decrease monotonically with temperature, reflecting the rapidly growing quark density as chiral symmetry is restored. The two relaxation times are nearly indistinguishable at low $T$ and converge further as $T$ increases. This behavior indicates that the scattering rate is dominated by the density of light quarks across the entire range of $T$ and $\mu_{B}$ studied, with a small contribution from the strange quarks. The simultaneous decrease in the relaxation times and constituent quark masses indicates that the high-temperature increase in the transport coefficients is governed by a competition between two effects: the growing thermal population of charge carriers and the progressively shorter microscopic relaxation time. The former increases the number of partons available for charge transport, while the latter limits the persistence of their momentum between successive collisions. The resulting diffusion coefficients therefore encode not only the abundance of thermally excited quarks but also the changing interaction rate of the medium. However, one thing to note here is that the relatively large relaxation times near the lower end of the temperature range indicate that the kinetic/hydrodynamic interpretation becomes less controlled there.\\

\begin{figure*}[ht!]
    \centering
    \includegraphics[width = 0.99\linewidth]{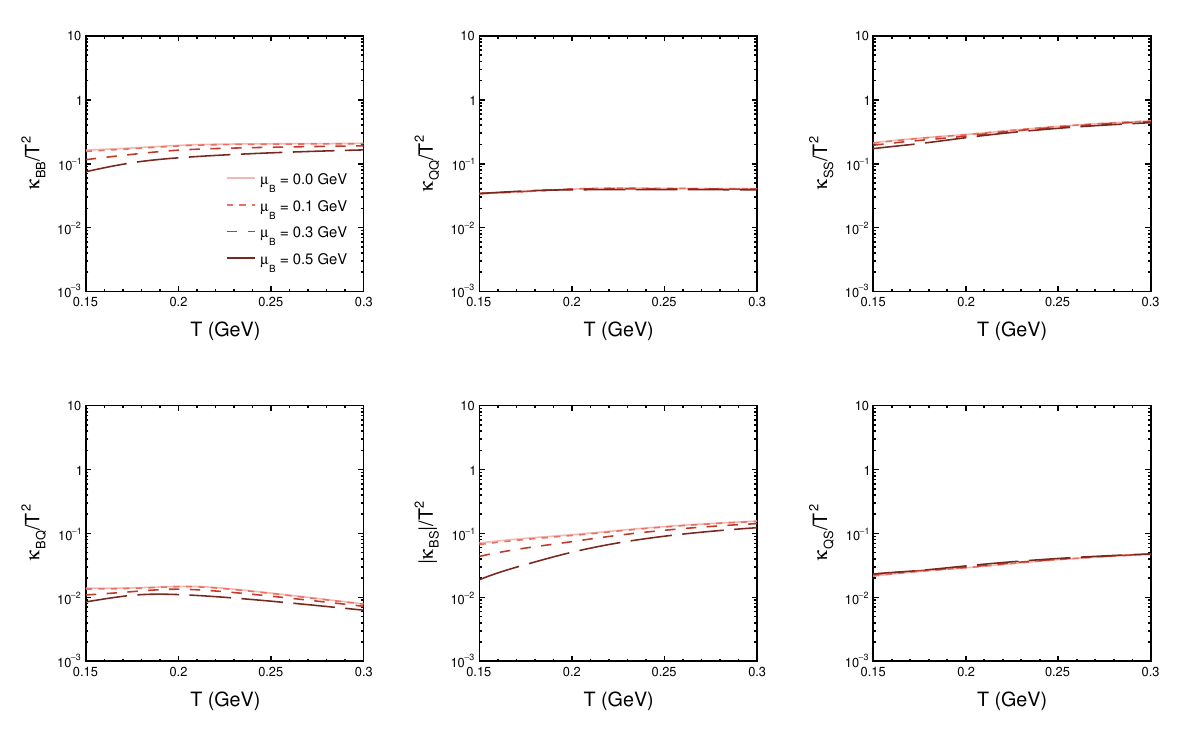}
    \caption{Diagonal and off-diagonal coefficients of the diffusion matrix as functions of temperature for different values of baryon chemical potential.}
    \label{Fig4}
\end{figure*}

Figure~\ref{Fig3} shows the diagonal ($\kappa_{BB}/T^{2}$, $\kappa_{QQ}/T^{2}$, $\kappa_{SS}/T^{2}$) and off-diagonal ($\kappa_{BQ}/T^{2}$, $\kappa_{BS}/T^{2}$, $\kappa_{QS}/T^{2}$) scaled diffusion coefficients at $\mu_B=0$, compared to the dynamical quasiparticle model (DQPM)~\cite{Fotakis:2021diq}, relativistic kinetic theory (Greif~\textit{et al.}~\cite{Greif:2017byw}), and holographic estimates~\cite{Rougemont:2017tlu}. The diagonal coefficients increase slightly with temperature, whereas the off-diagonal coefficients are nonzero throughout the temperature range studied. $\kappa_{BS}/T^2$ is negative and grows in magnitude with temperature, while $\kappa_{QS}/T^2$ is positive and of comparable magnitude. The coefficient $\kappa_{BQ}/T^2$, by contrast, decreases in magnitude with increasing temperature. We emphasize that a strictly vanishing $\kappa_{BQ}$ is only expected in the idealized limit of exactly massless, flavor-degenerate quarks at vanishing $\mu_{Q}$ and $\mu_{S}$.  Since $m_u \neq m_s$ at every temperature studied here, $\kappa_{BQ}$ remains finite throughout, providing a direct,
transport-level signature of residual $SU(3)_f$ breaking. It can be seen that our NJL results agree with other reported results qualitatively.\\

The simultaneous increase of the diagonal transport coefficients and the persistence of sizable off-diagonal components demonstrate that charge transport in the NJL medium cannot be interpreted as three independent diffusion processes. Rather, the transport response reflects the fact that the microscopic quark degrees of freedom carry multiple conserved quantum numbers simultaneously. The temperature dependence of the off-diagonal coefficients therefore contains information about how the relative contributions of the different flavor carriers evolve through the crossover region. In particular, the different temperature dependence of $\kappa_{BQ}$, $\kappa_{BS}$, and $\kappa_{QS}$ indicates that the flavor composition of the charge-carrying degrees of freedom is not simply encoded in the overall magnitude of the transport rate. Instead, the relative weighting of the light and strange sectors changes with temperature as the dynamically generated masses evolve. This makes the off-diagonal sector a useful transport observable for diagnosing flavor-dependent medium effects that are not visible from the diagonal coefficients alone. \\

\begin{figure*}[ht!]
    \centering
    \includegraphics[width = 0.99\linewidth]{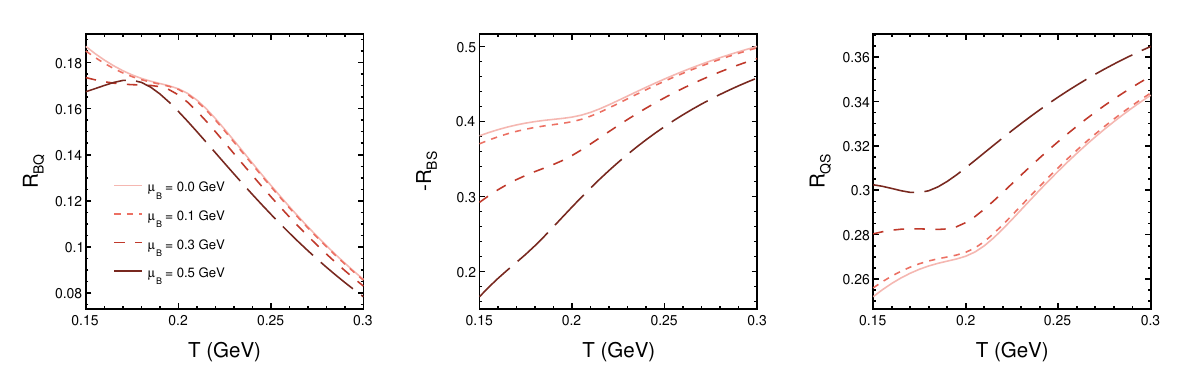}
    \caption{Normalized off-diagonal diffusion as a function of temperature for different values of baryon chemical potential.}
    \label{Fig5}
\end{figure*}

\begin{figure*}[ht!]
    \centering
\includegraphics[width = 0.99\linewidth]{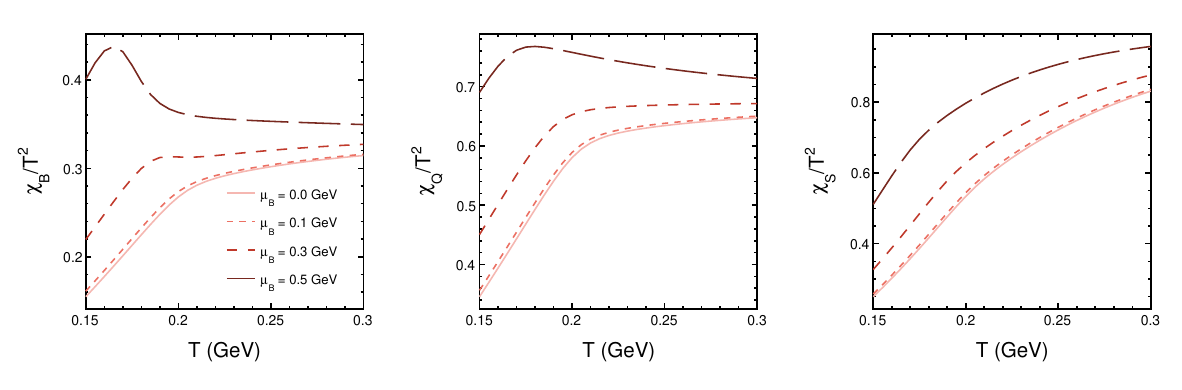}
    \ caption {Second-order susceptibility of baryon, electric charge, and strange quantum number as functions of temperature for different values of baryon chemical potential.}
    \label{Fig6}
\end{figure*}

Figure~\ref{Fig4} shows the diagonal diffusion coefficients estimated at finite baryon chemical potential in the NJL model. The baryon diffusion coefficient $\kappa_{BB}/T^2$ is suppressed with increasing $\mu_B$ at every temperature studied, but the size of the suppression is strongly temperature-dependent. This indicates that the sensitivity of baryon-number diffusion to the chemical potential is concentrated near the lower end of the temperature range studied, closer to the chiral crossover, and weakens substantially as the system moves toward the chirally restored regime. The suppression of $\kappa_{BB}$ with increasing $\mu_B$ indicates that increasing baryon chemical potential does not simply enhance baryon transport through the increased abundance of baryon charge carriers. Instead, the finite-density medium modifies the balance between charge-carrier phase space and their microscopic scattering dynamics. The strong temperature dependence of the $\mu_B$ effect further indicates that baryon-rich transport retains a pronounced sensitivity to the state of the medium near the crossover, whereas this sensitivity progressively weakens toward the chirally restored regime. A similar behavior is observed in $\kappa_{BS}/T^{2}$, reflecting the sensitivity of this coefficient to the strange-quark contribution to baryon and strangeness transport. In contrast, $\kappa_{QQ}/T^{2}$ and $\kappa_{QS}/T^{2}$ show almost no dependence on $\mu_{B}$. This reflects the fact that $\mu_{B}$ primarily controls the net baryon content of the medium, while electric charge and strangeness transport remain largely governed by the thermal population of their respective charge carriers. Since both light quarks and antiquarks contribute to electric charge transport, and strangeness is carried mainly by thermally excited strange quarks, these transport channels are comparatively less sensitive to changes in the baryon chemical potential. The behavior of $\kappa_{BQ}/T^{2}$ is slightly different, with the effect of $\mu_{B}$ remaining nearly uniform across the temperature range. This reflects the correlation between baryon-number and electric-charge transport, as increasing $\mu_B$ enhances the quark population relative to the antiquark population for all flavors, while their different electric charges determine how this baryon asymmetry contributes to the electric current. The nearly temperature-independent separation between the curves suggests that this baryon--charge correlation changes only weakly with temperature in the range studied.\\

To quantify the relative strength of the coupling between different conserved-charge sectors, we introduce the normalized off-diagonal diffusion coefficients,
\begin{equation}
R_{qq'}=
\frac{\kappa_{qq'}}
{\sqrt{\kappa_{qq}~\kappa_{q'q'}}},
\label{eq:Rqq'}
\end{equation}
where $(q,q')=(B,Q),(B,S),(Q,S)$. These dimensionless ratios are useful because the individual off-diagonal coefficients contain both the strength of the cross-charge coupling and the overall temperature dependence of the transport coefficients. By normalizing $\kappa_{qq'}$ to the corresponding diagonal scales, $R_{qq'}$ emphasizes the relative importance of cross-charge transport and allows the temperature evolution of the different coupling channels to be compared on a common scale. Figure~\ref{Fig5} shows the temperature dependence of $R_{BQ}$, $R_{BS}$, and $R_{QS}$ at finite $\mu_B$. The three coefficients exhibit clearly distinct behavior, demonstrating that cross-charge transport is sensitive to the microscopic composition of the charge-carrying degrees of freedom rather than simply to the overall strength of transport. The magnitude of $R_{BQ}$ decreases with increasing temperature, indicating a weakening sensitivity to the relative contributions of light and strange quarks as the flavor asymmetry of the medium is reduced. Quantitatively, $R_{BQ}\approx0.19$ at $T=0.15$ GeV and decreases to $\approx0.08$--$0.09$ at $T=0.30$ GeV, almost independently of $\mu_B$. Because $B$ and $Q$ are carried by the same quasiquarks in our approach, this is the cross-correlation expected for quark-carried baryon number; it provides a reference against which a different baryon carrier, such as the charge-neutral baryon junction invoked in Ref.~\cite{STAR:2024lvy}, could be tested. In contrast, $R_{BS}$ becomes increasingly negative while $R_{QS}$ increases, reflecting the growing contribution of thermally excited strange degrees of freedom that simultaneously carry baryon, electric, and strangeness quantum numbers. The opposite signs of the $BS$ and $QS$ correlations follow from the corresponding charge assignments of the strange quark.\\

\begin{figure*}[ht!]
    \centering
\includegraphics[width = 0.99\linewidth]{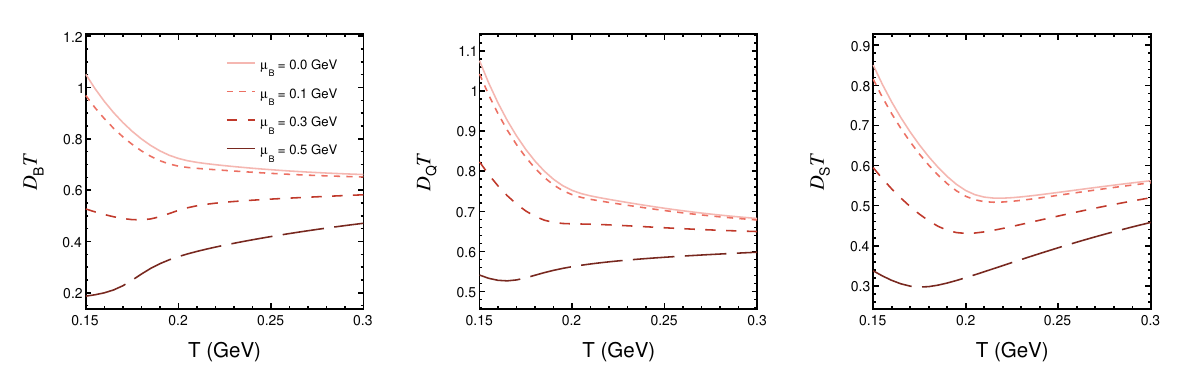}
    \caption{Diffusivity of baryon, electric charge, and strange quantum number as functions of temperature for different values of baryon chemical potential.}
    \label{Fig7}
\end{figure*}

\begin{figure*}[ht!]
    \centering
\includegraphics[width = 0.99\linewidth]{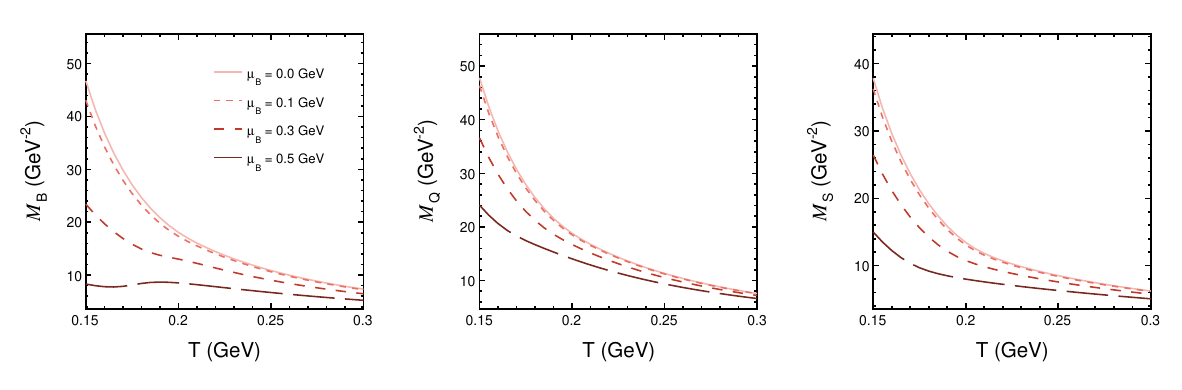}
    \caption{Mobility of baryon, electric charge, and strange quantum number as functions of temperature for different values of baryon chemical potential.}
    \label{Fig8}
\end{figure*}
The distinct evolution of the normalized coefficients therefore provides a transport-level view of how the relative composition of charge-carrying excitations changes with temperature. In particular, the weakening of the $BQ$ correlation together with the persistence and growth of the strange-sector correlations shows that different cross-charge channels respond differently to the evolution of the medium. The normalized off-diagonal coefficients consequently provide information complementary to the diagonal transport coefficients, helping to distinguish changes in the microscopic charge-carrier composition from changes in the overall efficiency of transport.\\

Figure~\ref{Fig6} shows the scaled conserved-charge susceptibilities $\chi_B/T^{2}$, $\chi_Q/T^{2}$, and $\chi_S/T^{2}$, estimated as full total derivatives $dn_q/d\mu_q$. All three susceptibilities increase monotonically with temperature at low $\mu_{B}$, consistent with the growing thermal quark population through the chiral crossover. The strangeness susceptibility exhibits the largest relative growth of the three, reflecting the continued, if slower, decrease of $m_s$ across the studied temperature range. The susceptibility results provide the thermodynamic counterpart to the transport behavior discussed above. The rapid increase of $\chi_q$ toward higher temperature implies that the medium develops an increasingly large capacity to accommodate fluctuations of the corresponding conserved charge. On the other hand, a strong $\mu_{B}$ dependence can be observed for all three susceptibilities, particularly the non-monotonic behaviour of  $\chi_B/T^{2}$ at low $T$ and high $\mu_{B}$. This enhanced sensitivity to $\mu_B$ indicates that charge fluctuations become increasingly constrained by the underlying nontrivial thermodynamic structure at finite density. Consequently, the interplay between susceptibilities and transport coefficients becomes particularly important for characterizing charge diffusion in the finite-density regime.\\

Figure~\ref{Fig7} shows the scaled diffusivities $D_B$, $D_Q$, and $D_S$, obtained from the Einstein relation from Eq.~\ref{eq2}. At vanishing and small $\mu_B$, the diffusivities generally decrease with increasing temperature. This inverted trend, relative to $\kappa_{qq}/T^2$, follows directly from the more rapid growth of the charge susceptibilities $\chi_q$ with temperature, which outpaces the growth of $\kappa_{qq}$ itself. At larger $\mu_{B}$, the diffusivities exhibit a minimum followed by an increasing trend. However, the dependence on $\mu_B$ becomes less pronounced as the temperature increases, and the curves tend to converge at higher $T$. The diffusivity is controlled by the competition between the microscopic transport rate and the thermodynamic susceptibility of the medium. Thus, the decrease of $D_q$ with increasing temperature reflects the fact that the rapid growth of the number of thermally accessible charge fluctuations can outweigh the increase of the underlying transport coefficient. 

Finally, Fig.~\ref{Fig8} shows the charge mobilities $M_B$, $M_Q$, and $M_S$, obtained from the Einstein--Smoluchowski relation. All three quantities generally decrease with increasing $T$ and baryon chemical potential. However, a subtle point to note here is that, at $T=0.150$~GeV, increasing $\mu_B$ suppresses $M_B$ far more strongly than $M_S$ and $M_Q$. As a result, $M_B$ and $M_S$ cross near intermediate $\mu_B$, while $M_Q$ remains the largest of the three throughout. The crossing of $M_B$ and $M_S$ at finite $\mu_B$ provides a nontrivial consequence of the coupled temperature and density dependence of transport and susceptibility. It demonstrates that the relative mobility of conserved charges is not determined solely by the microscopic abundance of their carriers; rather, it results from the competition between the corresponding transport coefficient and thermodynamic susceptibility. In particular, the stronger suppression of baryon mobility at finite $\mu_B$ implies that increasing baryon chemical potential modifies baryon-number transport more strongly than strangeness transport in this region of the phase diagram. The resulting change in the mobility hierarchy therefore identifies a density-dependent reorganization of the relative efficiency with which different conserved charges propagate through the medium. From a phenomenological perspective, this suggests that the relative relaxation of baryon, charge and strangeness fluctuations need not remain fixed across the phase diagram. Thus, net-baryon, net-charge and net-strangeness fluctuations may probe different transport timescales at different collision energies and baryon chemical potentials.\\

At small $\mu_B$, the baryon and charge mobilities are nearly equal, as expected when both charges are carried by the same quasiquarks; they separate only at large $\mu_B$, where baryon-density effects suppress $M_B$. Relative to this quark-carrier baseline, the enhanced baryon transport reported by STAR~\cite{STAR:2024lvy} would correspond to a larger baryon-to-charge response than obtained here.

Taken together, the results demonstrate that charge transport in the NJL medium is governed by a nontrivial interplay between microscopic scattering dynamics and thermodynamic charge fluctuations. The transport coefficients $\kappa_{qq}$ characterize the response of the medium to thermodynamic gradients, while the susceptibilities quantify the corresponding thermodynamic capacity for charge fluctuations; their ratio determines the diffusivity and therefore the efficiency of charge equilibration. The distinct behavior of $\kappa_{qq}$, $\chi_{q}$, and $D_{q}$, together with the finite-$\mu_{B}$ reordering of the conserved-charge mobilities, shows that no single transport quantity provides a complete characterization of charge propagation in the QCD medium.\\

\section{Summary and Outlook}
\label{summary}

We have studied the diffusion of the conserved charges $B$, $Q$, and $S$ in hot QCD matter within a three-flavor Nambu--Jona-Lasinio model in a kinetic-theory calculation of the diffusion matrix, susceptibilities, diffusivities, and conserved-charge mobilities. At $\mu_B=0$, the results were compared against dynamical quasiparticle, relativistic kinetic-theory, and holographic calculations and were then extended up to $\mu_B=0.5$~GeV. The off-diagonal coefficients provide information on the microscopic composition of the charge carriers: the $BQ$ channel is sensitive to the relative light- and strange-quark contributions, whereas $BS$ and $QS$ are dominated by strange degrees of freedom carrying multiple conserved charges. Their different temperature dependencies therefore encode changes in the flavor composition of the transport carriers beyond the overall transport strength. We further find that the generalized diffusivities decrease with temperature despite increasing diagonal transport coefficients, reflecting the competing growth of the charge susceptibilities. At finite $\mu_B$, the stronger suppression of baryon mobility relative to strangeness mobility produces a crossing of $M_B$ and $M_S$, while $M_Q$ remains the largest throughout, demonstrating that the relative efficiency of conserved-charge transport depends non-trivially on the thermodynamic state. The quantitative values of the transport coefficients remain subject to the phenomenological choice of the relaxation time in this study. A fully microscopic, flavor-dependent NJL scattering kernel is therefore needed to quantify the predicted mobility crossing. Since baryon number is carried by the quasiquarks in this approach, the results also serve as a quark-carrier reference for $B$--$Q$ transport, against which the recent STAR finding of a baryon-to-charge transport ratio above unity can be compared in future extensions. Future extensions to Polyakov-loop-improved NJL dynamics, the critical region, and a complete meson-exchange scattering treatment will allow the coupled diffusion modes and possible critical suppression of baryon diffusion to be studied more systematically.

\section*{Acknowledgement}
DS acknowledges the support from the postdoctoral fellowship of the DGAPA UNAM. K.G. acknowledges financial support from the Prime Minister’s Research Fellowship (PMRF), Government of India. R.S. acknowledges financial support from the DAE-DST, Government of India, under the mega-science project “Indian participation in the ALICE experiment at CERN” (Project No. SR/MF/PS-02/2021-IITI(E-37123)).

\end{document}